# Mobile Based Gradebook with Student Outcomes Analytics

[1]Ronel B. Dayanghirang, [2]Alexander A. Hernandez

Technological Institute of the Philippines Manila, Philippines
Email: [1]roneldayanghirang@gmail.com, [2]alexander.hernandez@tip.edu.ph
Contact:[1]+63-9053140315, [2]+639154860445

*Abstract* - Mobile applications and other integration of information and communication technology (ICT) have become well-known in education to monitor teaching and learning activities. The analysis of student learning through evaluation is a growing area of interest for teachers in higher education aiming to enhance students learning experience. This paper describes a development of student outcomes monitoring tool that applies analytics to provide feedback to students as they progress in the ladder of achieving the intended learning outcomes. The student outcomes focus on the core elements of the curriculum; it offers detailed student outcomes where the result in courses evaluations and recordings are tracked and analyzed. The data revealed that the student outcomes monitoring and analytics tool is adequate in providing constant feedback to students on the achievement of the desired learning outcomes as well as support teachers in planning the teaching and learning activities, enhance feedback system, academic planning and improvement.

## I. INTRODUCTION

Advanced education establishments are progressively offering units in online and mixed conveyance modes. In any case, the regular heuristics that staff depends upon to distinguish separation are not promptly transferable to or accessible in, the online connection. The decreased contact and quickness makes it more difficult for them to be aware of how their students are engaging [1].

The universality of grading management systems implies that numerous connections between students, peers, instructors, and content are captured in databases. The youthful field of learning analytics (and the firmly adjusted field of educational information mining) seeks to make sense of these and other information to understand better and advance student learning [2].

The greater part of work in learning analytics to date has focused on improving student performance and retention by determining variables that are indicative of issues in these areas [3]. To close the analytics loop and sanction change, student information should be properly comprehended and acted upon [4]. To this end, some staff-confronting dashboards that graphically speak to student information have been conceptualized and developed [5] [6]. These typically seek to assist in deciphering complex student interactions and provide information for analyzing processes about learning and teaching [7]. Such choices may include activating and sending intercessions, encouraged by frameworks that allow staff to contact students and give auspicious counsel and input[8] [9].

Thus, there is a need to provide a system to enhance the process of grading and evaluating of the students. And creating a mobile-based gradebook with student outcomes analytics can assist teachers in measuring and monitoring student learning skills and evaluation of what they have learned. Mobile technology would help them to learn and ease their work. This project aims to present a mobile based gradebook and a system that create an outcomes analytics based on students' performance. The outcomes focus on the core elements of the curriculum, ensuring that every learning activity, and providing opportunities for students to demonstrate proficiency in a variety of modalities. Hence, students will have a clear view of what the teacher expect for success and being prepared to show what they know and to reach their learning targets and competencies.

## II. RELATED WORKS

Various higher education institutions are beginning to utilize learning analytics to help students and staffs comprehend and upgrade knowledge. Some new Office of Learning and Teaching projects have focused on constructing institutional frameworks around advancing learning analytics, analyzing data from social media interactions and understanding how data can be used by teachers [10][11].

As the case in any new area of practice and research, for example in analytics, many terms have been presented that have conflicting useful or theoretical definitions. For sure, the term analytics holds different meanings for different people [3]. Learning analytics is the estimation, gathering, analysis, and detailing of information about learners and their unique circumstances, with the end goal of comprehension and - improving learning and the environments in which it occurs [12]. These intelligent information frameworks can be utilized to enhance instructing and learning as a significant aspect of a procedure that is learner-delivered, in nearness to the learning occasion. It can help to discover and reveal information and make connections of a course or program level that can be utilized to create predictive models and can be used for scholarly analytics, which is the utilization of business knowledge in education and exists at an institutional, local, and national/worldwide level [7].

Past learning and scholastic analytics, Campbell, DeBlois, and Oblinger anticipate that the present patterns uncover the graduation rates amid this present data age are on course to reveal instructive holes, and a two percent decrease in higher education's [13]. While a few differences might be made in definitions between learning analytics and scholarly analytics, in this paper, these terms are utilized nearly synonymously.

As a developing field, learning analytics depends on information winnowed from different sources to settle on choices about academic progress, forecasts about future performance, and to recognize potential issues [14]. Among the most significant challenges facing distance education has been the lack of knowledge about the ways that students interact with learning materials. Large portions of the attributes of distance education that make it an appealing option for students, educators, and administrators are similar components that make analyzing and assessing course and program adequacy a challenge. Since students are at-a-remove, educators don't get similar sorts of input from students (explicit or implicit) that they get in a customary eye to eye classroom.

## III. METHODOLOGY

This project uses Agile Software Development method to ensure that all user specification, development, and outcomes are met. This study uses PHP, JQuery, JavaScript programming and third party technologies. This study includes different user, Department Head, Instructors and students. This research used a software evaluation following ISO9126 criteria and distributed to the users during the testing stage. The Likert's scale with the interpretation of highly acceptable, acceptable, moderately acceptable, slightly acceptable, not acceptable was used to specify the users' level of agreement or disagreement on the software evaluation items. The results of software evaluation are presented in the succeeding sections of this paper.

## IV. RESULTS AND DISCUSSION

The following parts present the mobile-based gradebook with student outcomes analytics. Figure 1 presents the system architecture in both web and mobile platform covering account management, grade management, and analytics while the web modules include subject management, classes data entry, evaluation data entry, settings, and reports. These modules are intended for the instructors and department head of the institution.

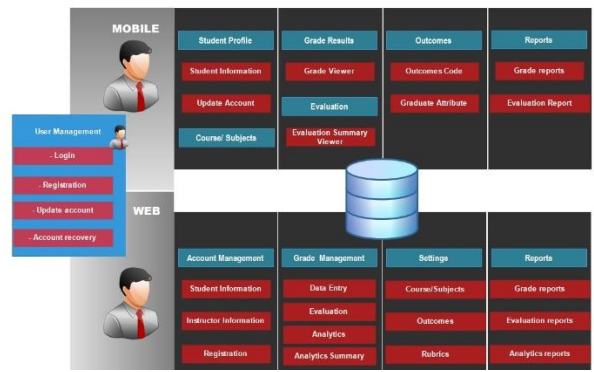

Figure 1 Mobile based Gradebook with Student Outcomes Analytics Architecture

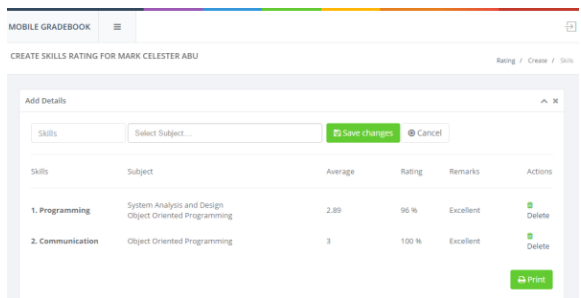

Figure 2 Skills Rating

The skills rating module comprises searching of student skills information and evaluation details. This feature also present the different skills of the student based on their evaluation per outcomes. Overall, this feature allows the student to view his rating information on a particular skill. This skills rating feature has two part the skills textbox and the course dropdown which enable the user to create skills score in selected courses for skills monitoring purpose. This skills rating feature make users' focus on the core outcomes they want to improve.

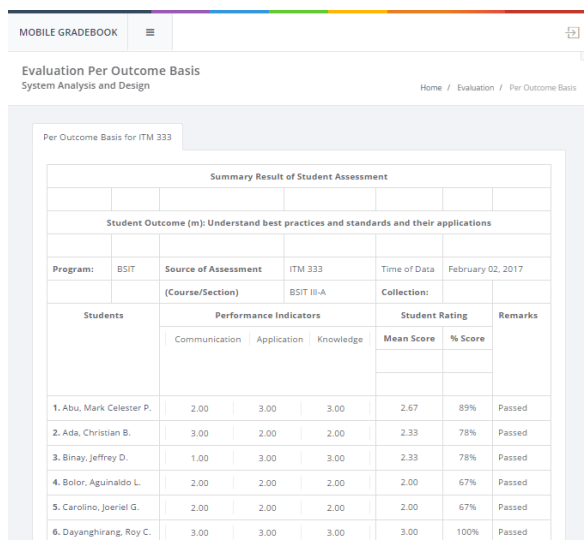

Figure 3 Evaluation Module

Figure 3 presents Evaluation feature were the teachers evaluate student thru the help of outcomes rubrics. This system module efficiently shows the results of the evaluations. The data serve as a guide or indicator of the effectiveness of their program and help them to determine the current student learning status. Thus, it can help the teachers to determine the student learning experience in every student's outcomes. This gives user to know how things are working, to know which parts of programs are working the best, and which student outcomes need to be change or improve.

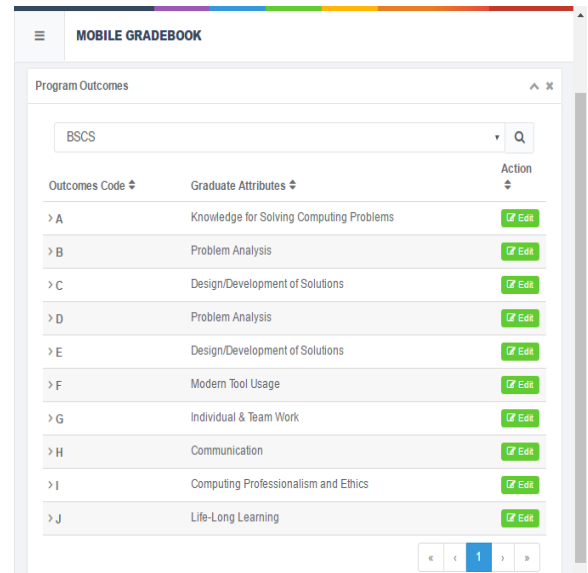

4 Program Outcomes

Figure 4 shows Program Outcomes module. It presents the outcomes code and the graduate attributes. This feature allows creating program outcomes based on the curriculum and this program outcomes features allows updating the curriculum and making it easier to know the curriculum that is currently in use. Also, this feature is an easy way in were the student and instructors can view every learning outcomes and what is the graduate attribute meaning of every outcomes code.

The contribution to course planning is it can help them determine what is working and what is not working in their courses or programs and user can provide powerful evidence based on the data to justify needed resources to maintain or improve programs specially allow user to make improvements in the program outcomes.

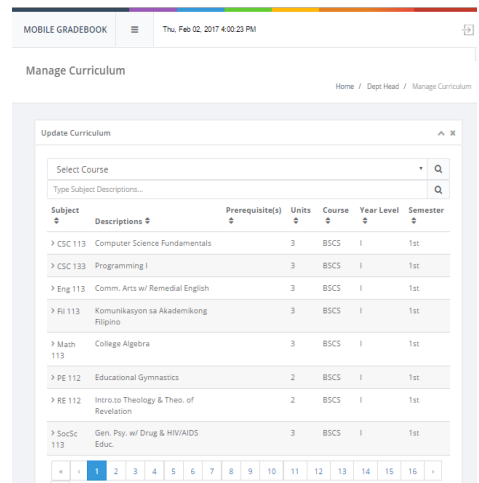

Figure 5 Curriculum Module

Figure 5 shows the curriculum features. This feature contains course offered by the school, and it is manageable base on what curriculum are currently offered by the school. This feature presents the capability for the user to quickly view what the offered courses are, how many units are the courses have and their pre-requisites.

Figure 6 Course Mapping Module

This feature offers the capability for the Program Chairperson to track the outcomes of every course. The module contains courses that have outcomes and can be updated based on the Commission on Higher Education Memorandum Order. In this feature, outcomes in every subject can be tick (the star logo will appear) to enabled it and unticked to disabled. And it is recorded it also add to the course mapping information to give the teacher an easy way to know if there's an outcome in a particular course. This gives useful data that help administrators to improve academic planning and decision-making and to make improvements in student outcomes.

Figure 7 Analytics Module

The mobile based gradebook with student outcomes analytics also provide a feature for analytics reports (Figure 7). This module efficiently shows the analytics report of the student based on their evaluation it shows the tallies of how many students got the same score in a particular outcome. This feature student's information stored in the mobile application database systems.

Figure 8 Analytics Report in Bar Graph

This feature offers the capability for the student to view analytics report thru graph to easily track what outcomes he/she need to focus more (Figure 8- 9). It also helps the user to know the percentage of how many students got the same score in a particular outcome. The contribution of this to course planning is; it is providing valuable data to monitor student outcomes and when the administrators see the results it allows them to make better decisions about programs andmake an institutional commitment to continually improve the academic programs and services.

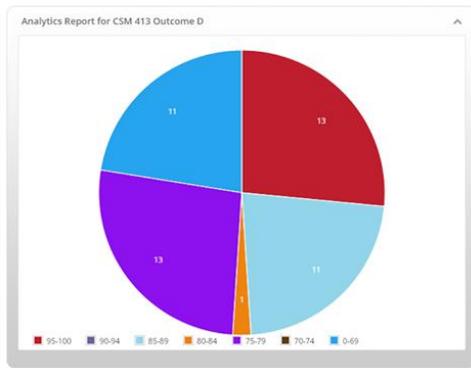

Figure 9 Analytics Report in Pie Graph

Table 1 Summary of Software Evaluation

| Criteria | Mean | Interpretation |
| --- | --- | --- |
| 1. Functionality | 4.49 | Acceptable |
| 2. Reliability | 4.53 | Highly Acceptable |
| 3. Usability | 4.34 | Acceptable |
| 4. Efficiency | 4.49 | Acceptable |
| 5. Maintainability | 4.52 | Highly Acceptable |
| 6. Portability | 4.20 | Acceptable |
| **Overall Weighted Mean** | **4.43** | **Acceptable** |

In summary, the software evaluation indicates a strong perception among the respondents that the Mobile based gradebook with student outcomes analytics is functional (4.49), highly reliable (4.53), usable (4.34), efficient (4.49), highly maintainable (4.52), and portable (4.20). Hence, the software evaluation receives an overall rating of 4.43 with an interpretation of acceptable. The results indicate that the mobile application performance has certainly achieved the goal of the study of creating a mobile based gradebook with student outcomes analytics to enhance and improve the manual process of grading and evaluating of the students of computer studies department.

**V. CONCLUSION AND RECOMMENDATIONS**

This research aims to provide a mobile based gradebook with student outcomes analytics specifically develop a tool to monitor the achievement of student outcomes and applies analytics. The software evaluation also indicates the relevance of the features provided for instructors and students to ensure that information is shared efficiently. However, this research has also recommendations to enhance the findings of the study including; (a) utilize the program by means of providing more learning outcomes based on Commission on Higher EducationMemorandum Order; (b) improve the update options for gradebook and evaluations to create different graphics results in student outcomes analytics.; (c) explore the perceptions of students in the use of mobile based gradebook with student outcomes analytics; (d) integration of mobile based gradebook with student outcomes analytics to college enrollment system for more accurate data.

**References**

[1] Richardson, J., & Swan, K. (2003). Examing social presence in online courses in relation to students' perceived learning and satisfaction.

[2] Siemens, G., & d Baker, R. S. (2012, April). Learning analytics and educational data mining: towards communication and collaboration. In *Proceedings of the 2nd international conference on learning analytics and knowledge* ACM.

[3] Romero, C., & Ventura, S. (2013). Data mining in education. *Wiley Interdisciplinary Reviews: Data Mining and Knowledge Discovery*, *3*(1), 12-27.

[4] Clow, D. (2012, April). The learning analytics cycle: closing the loop effectively. In *Proceedings of the 2nd international conference on learning analytics and knowledge* (pp. 134-138). ACM.

[5] Duval, E. (2011, February). Attention please!: learning analytics for visualization and recommendation. In *Proceedings of the 1st International Conference on Learning Analytics and Knowledge* (pp. 9-17). ACM.

[6] Pardo, A. (2014). Designing learning analytics experiences. In *Learning Analytics* (pp. 15-38). Springer New York.

[7] Siemens, G., & Long, P. (2011). Penetrating the Fog: Analytics in Learning and Education. *EDUCAUSE review*, *46*(5), 30.

[8] Mattingly, K. D., Rice, M. C., & Berge, Z. L. (2012). Learning analytics as a tool for closing the assessment loop in higher education. *Knowledge Management & E-Learning: An International Journal (KM&EL)*, *4*(3), 236-247.

[9] Atif, A., Froissard, C., Liu, D., & Richards, D. (2015). Validating the effectiveness of the Moodle Engagement Analytics Plugin to predict student academic performance.

[10]   Kitto, K., Cross, S., Waters, Z., & Lupton, M. (2015, March). Learning analytics beyond the LMS: the connected learning analytics toolkit.

[11]   Kennedy, G., Corrin, L., Lockyer, L., Dawson, S., Williams, D., Mulder, R., ...& Copeland, S. (2014). Completing the loop: returning learning analytics to teachers.

[12]   Shum, S. B., & Ferguson, R. (2012). Social learning analytics. Educational technology & society, 15(3), 3-26.

[13]   Campbell, J. P., &Oblinger, D. G. (2007). Academic analytics. *Educause Quarterly*.

[14]   Johnson, L., Adams, S., & Cummins, M. (2012). Technology Outlook for Australian Tertiary Education 2012-2017: An NMC Horizon Report Regional Analysis. New Media Consortium. 6101 West Courtyard Drive Building One Suite 100, Austin, TX 78730.